\documentclass[prl,twocolumn,showpacs,nofootinbib]{revtex4}

\usepackage[dvips]{graphicx}
\usepackage{amsmath}
\usepackage{amssymb}
\usepackage{color}

\newcommand{\red}[1]{\textcolor{black}{ #1 }}

\begin{document}

\preprint{LA-UR-05-3234}

\title{Pinning frequencies of the collective modes in $\alpha $-uranium}
\author {B.~Mihaila, C. P.~Opeil, F. R.~Drymiotis, J. L.~Smith, J. C.~Cooley,
         M. E.~Manley, A.~Migliori, C.~Mielke, T.~Lookman, A.~Saxena, A. R.~Bishop,
         K. B.~Blagoev, D.~J.~Thoma, and J.~C.~Lashley}
         \affiliation{Los Alamos National Laboratory, Los
Alamos, New Mexico 87545, USA}

\author {B.E.~Lang, J.~Boerio-Goates, and B.F.~Woodfield}
\affiliation{Department of Chemistry and Biochemistry, Brigham Young
Universtiy, Provo, Utah 84602, USA}

\author {G.M.~Schmiedeshoff}
\affiliation{Department of Physics, Occidental College, Los Angeles, California 90041, USA}

\begin{abstract}
Uranium is the only known element that features a charge-density
wave (CDW) and superconductivity. We report a comparison of the
specific heat of single-crystal and polycrystalline
$\alpha$-uranium. \red{Away from the the phase transition the
specific heat of the polycrystal is larger than that of the single
crystal, and the aim of this paper is to explain this difference.}
In the single crystal we find excess contributions to the heat
capacity at 41~K, 38~K, and 23~K, with a Debye temperature,
$\Theta_D$ = 256~K. In the polycrystalline sample the heat capacity
curve is thermally broadened ($\Theta_D$ = 184~K), but no excess
heat capacity was observed. The excess heat capacity, $C_\phi$
(taken as the difference between the single crystal and polycrystal
heat capacities) is well described in terms of collective-mode
excitations above their respective pinning frequencies. This
attribution is represented by a modified Debye spectrum with two
cutoff frequencies, a pinning frequency, $\nu_o$, for the pinned CDW
(due to grain boundaries in the polycrystal), and a normal Debye
acoustic frequency occurring in the single crystal. We explain the
50-year-old difference in Debye temperatures between heat capacity
and ultrasonic measurements.
\end{abstract}

\maketitle

New ground states and modulated structures arise in low-dimensional
solids~\cite{Peierls}\red{, but are of particular interest when they
occur in solids that are not low-dimensional, such as
$\alpha$-uranium ($\alpha$-U)~\cite{over}}. When periodic, these
ordered structures may be a spin-density wave (SDW) as found in
chromium~\cite{steve} or a charge-density wave (CDW) as found in
metal dichalcogenides~\cite{Gruner1} and in
$\alpha$-U~\cite{Lander}. Uranium is unique in the periodic table in
that it is the \emph{only} element known to exhibit a CDW \red{in
the absence of distortions due to a SDW.} Although detailed
band-structure calculations on the CDW transitions have been carried
out~\cite{Lars}, knowledge of the CDW energetics in $\alpha$-U
remains to be experimentally verified.

Across the light actinide elements (thorium -- plutonium) itinerant
5f-electrons lower their energies by causing Peierls-like
distortions~\cite{peierls_foot}, leading to the lowest-symmetry
structures in the periodic table. Like their quasi-one- and
two-dimensional counterparts, the collective mode in the crystal is
characterized by an energy gap in the single-particle excitation
spectrum originating from a charge modulation of the periodic
lattice~\cite{Gruner2},
$
   \rho(x)
   \ = \
   \rho _o
   \ + \
   \Delta \rho \ \cos ( 2k_F x  + \phi )
$, where $\rho_o$ is the density of the normal state, $\Delta \rho$
is the amplitude of the CDW, k$_F$ is the Fermi wave vector, and
$\phi$ is a phase factor defining the position of the periodic
modulation.

In a perfect lattice, a CDW should be able to slide throughout the
crystal without resistance like a superconductor, as suggested by
Fr\"ohlich~\cite{frohlich}. However, in real crystals pinning by
disorder from microstructure (defects, twins, grain boundaries,
impurities, surfaces, etc.) elevates the collective mode to finite
frequencies expressed by a threshold electric field,
$E_{\mathrm{th}}$. The threshold field is usefully characterized
within a single-particle model~\cite{Gruner1}. In this model, the
collective mode is described by the equation of motion of a
classical particle moving in a periodic potential. If a small dc
field is applied near the threshold field, the particle may be
displaced from the bottom of the well. For fields less than
$E_{\mathrm{th}}$ the particle remains localized within the
potential. For electric fields greater than $E_{\mathrm{th}}$ the
particle slides down the potential leading to an extra contribution
to the conductivity generated by motion of the collective mode
through the crystal, or Fr\"ohlich conduction.

Significant differences in single and polycrystalline uranium have
been observed in the thermal expansion, high temperature neutron
diffraction and calorimetry~\cite{Jousset,manley}. The temperature
dependence of the thermal expansion coefficients for single crystal
uranium differ in the three orthorhombic principle
directions~\cite{Lander}. Presumably a simple ensemble average for a
polycrystal sample would display a temperature dependence different
from any in the single crystal. Therefore it was anticipated that a
comparison of the single- and polycrystalline specific heat would
provide a useful probe of the effects on the Debye spectrum
associated with microstructural effects. In this Letter we show that
the differences between the specific heat of crystals that undergo
the weakly pinned CDW transitions (single crystals) and those that
are strongly pinned CDW (polycrystals) provide a definitive way to
determine the energetics of pinning.

Biljakovic \emph{et al.}~\cite{monceau} have described the effect of
collective mode dynamics on the phonon spectrum. In these cases, a
CDW modifies the phonon spectrum, which can be described by a
mixture of acoustic and optic modes. The modified Debye spectrum
features two cutoff frequencies: a lower frequency corresponding to
a pinned state, $h \nu_o = k_B T_o$, and an upper frequency, $h
\nu_\phi = k_B \Theta_\phi$, corresponding to the normal Debye
temperature for the collective modes~\cite{monceau}. With this
modification, the excess heat capacity $C_\phi$ (i.e. the difference
between the specific heats of the single- and polycrystal, $C_\phi =
C_\mathrm{single}-C_\mathrm{poly}$)
can be written as~\cite{excess_heat}
\begin{equation}
   C_\phi^\mathrm{fit} =
   3 N_\phi k_B \,
   \left( \frac{T}{\Theta_\phi}
   \right)^3
   \int\limits_{
        \raise0.2ex\hbox{$T_o$} \!
        \mathord{\left/{\vphantom {{T_o } T}}\right.
        \kern-\nulldelimiterspace}\!
        \lower0.2ex\hbox{$T$}}^{
        \raise0.2ex\hbox{$\Theta_\phi$}\!
        \mathord{\left/{\vphantom {{\Theta_\phi  } T}}\right.
        \kern-\nulldelimiterspace}\!
        \lower0.2ex\hbox{$T$}}
   \mathrm{d}x \,
   \bigl( x - x_o \bigr)^2 \,
   \frac{x^2 \ e^x }
        {\bigl( {e^x  - 1} \bigr)^2}
   \>,
\label{eq:cphi}
\end{equation}
where we have introduced the notation $x = h\nu/(k_B T)$, with k$_B$
being the Boltzmann constant, and $N_\phi$ denotes the number of
excitations involved in the transition.

Single crystals of uranium were formed from electro-transport in a
molten LiCl-KCl eutectic electrolyte containing UCl$_3$ at
approximately 3 wt~\%, as described elsewhere~\cite{jom}. After the
single crystals were measured, they were cast through induction
melting of the single crystals in a BeO crucible, under an inert
atmosphere, to prepare polycrystalline samples of the same pedigree.
The sample was melted only once to minimize the possibility of
contamination from the crucible or the carrier gas, then cleaned in
concentrated nitric acid. The grain sizes were 20-50 microns. The
low-temperature specific heat of the samples were measured using a
semi-adiabatic calorimeter~\cite{lash1} in the temperature range
\red{$\sim$0.5} to 100~K. The single crystal had a mass of 0.5 g and
the polycrystalline sample mass was 1.4 g.

\begin{figure}[t!]
\includegraphics[width=\columnwidth]{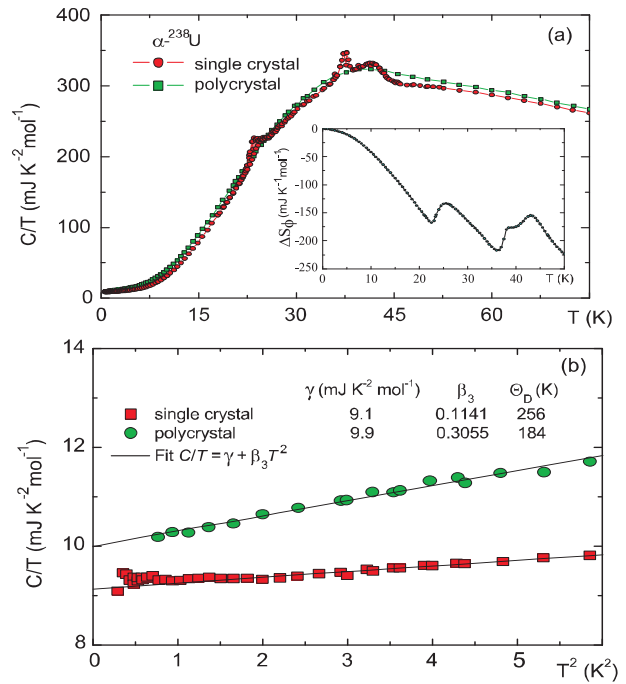}
\caption{(Color online) Measured specific heat plotted as $C/T$ vs.
$T$. The transitions at 38 and 41~K are broadened in the
polycrystal, as the specific heat for the polycrystal rises more
quickly before the onset of the transition in the single crystal.
The inset shows the temperature dependence of the excess entropy
\red{$\Delta S_\phi = S_\mathrm{single}-S_\mathrm{poly}.$}}
\label{Figure1}
\end{figure}

The effect of microstructure on collective mode pinning is shown in
Fig.~\ref{Figure1}. The C/T versus T curves for polycrystal uranium
and single crystal show a different contribution throughout the
temperature range encompassing the CDW state. Specifically, three
CDW transitions are observed in the single crystal at 23~K
($\alpha_3$), 38~K ($\alpha_2$), and 41~K ($\alpha_1$). While the
polycrystalline sample shows no sharp excess heat capacity in Fig.
1, one sees a significant difference in the \red{single-crystal}
sample. Starting below the $\alpha_3$ transition, the heat capacity
of the polycrystalline sample is larger. The heat capacity is lower
at the $\alpha_3$ transition returning to a larger value upon
warming to temperatures between the $\alpha_3$ and $\alpha_2$
transitions. Similarly it returns to a lower value through
$\alpha_2$ and $\alpha_1$ transitions. This excess at the
transitions is shown in the excess entropy, \red{$\Delta S_\phi =
S_\mathrm{single}-S_\mathrm{poly}$}, depicted in the inset of
Fig.~\ref{Figure1}a. The excess contributions are consistent with
the formation of an energy gap in the single crystal that  is much
larger than that observed in the polycrystal. Above the CDW
transitions the difference in entropy has been determined through
neutron diffraction and calorimetric measurements of Manley and
coworkers to arise from a microstrain contribution at the grain
boundary interfaces in polycrystalline uranium~\cite{manley}.

As with familiar 1D and 2D CDWs, we anticipate that the CDW
transitions correspond to a lock-in commensurate with the
periodicity of the underlying lattice, \red{separated} by mixed
phases of commensurate spatial regions and
discommensurations~\cite{bak}. Phonon softening in $\alpha$-U has
been observed below 50~K~\cite{smith}. The new unit cell doubles in
the [100] direction ($a$~direction) at T$_{\alpha_1}$ = 41~K,
becomes six times larger in the [010] direction ($b$~direction) at
T$_{\alpha_2}$ = 38~K, and becomes nearly six times larger in the
[001] direction with the same periodicity (commensurate) along the
$b$~direction, at T$_{\alpha_3}$ = 23~K~\cite{Lander}. Consequently,
the volume of the new cell is 6000 \AA$^3$. Pinning of the
collective modes by microstructure, also consistent with thermal
expansion measurements, results in the polycrystal exhibiting a
minimum in ($\Delta$L/L) between 45 and 50~K. The relative volume
change between 4 and 50~K is significantly larger in single
crystals~\cite{Lander,Jousset}.

Further evidence for the collective-mode pinning is provided by the
linear region of the low-temperature specific heat (see
Fig.~\ref{Figure1}b), where we notice a pronounced difference in the
low-temperature Debye limit. Figure~\ref{Figure1}b shows an expanded
view of $C/T$ vs.~T$^2$. Below 10~K, each data set was fit to the
electronic \red{and lattice} specific heat
\begin{equation}
   \raise0.2ex\hbox{$C$} \!\mathord{\left/{\vphantom {C T}}
   \right.\kern-\nulldelimiterspace}\!\lower0.2ex\hbox{$T$}
   \ = \ \gamma \ + \ \beta_3 \ T^2
   \>.
\label{eq:covert}
\end{equation}
The Debye temperature, $\Theta_D$, was obtained from $\beta_3$, as
\begin{equation}
   \Theta_D
   =
   \sqrt[3]{\raise0.2ex\hbox{$12\pi ^4R$}\!\mathord
   {\left/{\vphantom {{12\pi ^4 R} {5\beta_3}}}\right.\kern-\nulldelimiterspace}
   \!\lower0.2ex\hbox{$5\beta_3$}}
   \>,
\label{eq:debyet}
\end{equation}
while the $y$-intercept gives the electronic specific
heat,~$\gamma$. The parameters obtained from the fit to
Eq.~\eqref{eq:covert}, and the inferred Debye temperatures are
listed in Fig.~\ref{Figure1}b. The most striking difference between
the two samples is that the slope, $\beta_3$, increases by a factor
of~3 in the polycrystalline sample, resulting in a 35\% reduction
in~$\Theta_D$~\cite{lambda}.

\begin{figure}[t]
\includegraphics[width=\columnwidth]{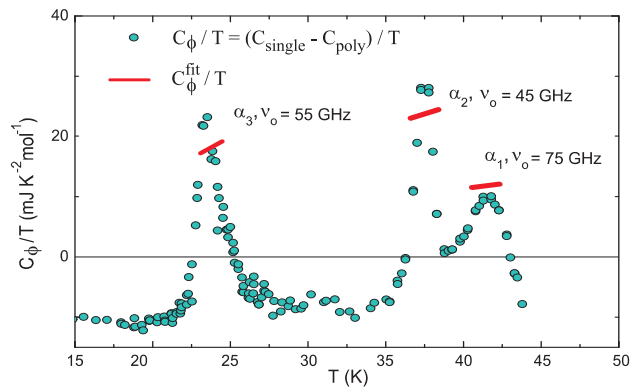}
\caption{(Color online) Excess specific heat \red{plotted as
$C_\phi/T$ vs. $T$.} Note, the excess specific heat can be 
negative since we neglected the pinning of the collective mode in
the single crystal. The lines depict \red{$C_\phi^\mathrm{fit}/T$ in
the peak region, where $C_\phi^\mathrm{fit}$} is calculated from the
modified Debye law, Eq.~\eqref{eq:cphi}, with the individual pinning
frequencies shown above each transition.} \label{Figure2}
\end{figure}

The large differences in the specific heat of the single-crystal and
polycrystalline samples between 15~K and 60~K (Fig.~\ref{Figure1}a)
and the low-temperature limiting Debye temperatures
(Fig.~\ref{Figure1}b) suggest a modified Debye spectrum, as in
Eq.~\eqref{eq:cphi}. The excess heat capacity, $C_\phi$, as a
function of temperature is depicted in Fig.~\ref{Figure2}. In other
words $C_\phi$ represents the heat capacity of the collective modes
above their respective pinning frequencies. Anomalies such as these
are known to arise from the phonon-branch splitting into an optic
mode and an acoustic mode below the CDW condensate formation. Such
features cannot be reproduced from a normal Debye model or
Debye/Einstein models, and in order to analyze the three anomalies,
it is necessary to consider the two different Debye temperatures
obtained in Eq.~\eqref{eq:debyet} and to fit the excess specific
heat, C$_\phi$, using the integral evaluated at temperatures near
the transitions. This description of the modified Debye spectrum is
further evident from the fact that $\Theta_D$ obtained from the
single crystal is the only calorimetric value~\cite{lash1} to match
the value obtained from elastic constant measurements~\cite{Ed}.

\begin{figure}[b]
\includegraphics[width=\columnwidth]{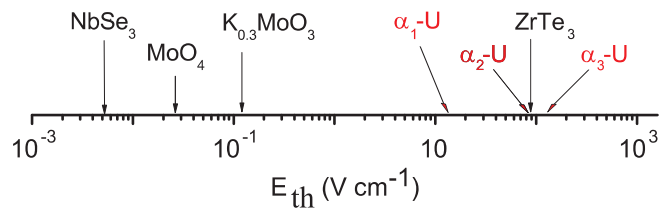}
\caption{(Color online) Estimated electric field threshold energies
for $\alpha_1$ and $\alpha_2$ transitions in $\alpha$-U shown with
some common quasi-one and two dimensional CDW systems.}
\label{Figure3}
\end{figure}

\red{Keeping N$_\phi$ and $\nu_o$ as free parameters, and
$\Theta_\phi$ fixed to the single-crystal} value listed in
Fig.~\ref{Figure1}b
, the low-temperature fit for the polycrystalline sample leads to
the following $N_\phi/N_A$ values: 0.25 for $\alpha_1$, 0.55 for
$\alpha_2$, and 0.9 for $\alpha_3$ (N$_A$ is the Avogadro number).
The values obtained for the pinning frequencies are indicated in
Fig.~\ref{Figure2} for each transition~\cite{footnote}. These values
are consistent with electronic structure calculations~\cite{Lars},
but pinning to microstructure at $\alpha_1$ occurs by the opening of
two gaps in the Fermi surface, accompanied by a doubling of the unit
cell in the $a$~direction. In essence, the two areas of the Fermi
surface with the highest density of states are nested in the
$a$~direction. Because there are two energy scales involved in this
transition (the lattice distortion energy and the electronic energy)
and because there is a nesting in the $b$~direction only 2~K lower
than the $\alpha_1$ transition, it is apparent that the $\alpha_2$
transition is necessary to overcome the overall lattice distortion
energy: the doubling of the unit cell in the $a$~direction at 41~K
leaves the Fermi surface in a favorable geometry for nesting in the
$b$ direction, which is possible because there is elastic energy
available to further lower the electronic energy (see inset in
Fig.~\ref{Figure1}a).

The estimated values of the pinning frequency are lower than those
obtained for NbTe$_4$ (540~GHz) and KCP (200~GHz) but similar to
(TaSe$_4$)$_2$I (42~GHz), where nonlinear conduction due to CDW
motion is observed~\cite{mori}. From the pinning frequencies, we
estimate a lower bound on the threshold electric
field~\cite{Gruner2}, as
\begin{equation}
   E_{\mathrm{th}} \ = \ \frac{1}{2} \
   \raise0.2ex\hbox{$m^* \omega_o^2$}\!
   \mathord{\left/{\vphantom {{m^* \nu_o^2 } {k_F e}}}\right.
   \kern-\nulldelimiterspace}\!
   \lower0.2ex\hbox{${e\,k_F}$}
   \>,
\end{equation}
where m$^*$ is the effective mass and $\nu_o=\omega_o/(2\pi)$ is the
pinning frequency obtained from the fit of the excess specific heat
capacity to Eq.~\eqref{eq:cphi}. With m$^*\,$k$_F$ = 16.6 $\times$
10$^{-30}$~eV~s$^2$/atom, one obtains $E_{\mathrm{th}}$ as 14.5, 89,
and 133~V/cm for the $\alpha_1$, $\alpha_2$, and $\alpha_3$
transitions, respectively. Here, we have approximated the period
($\pi/k_F$) of the CDW, by 2$a$ in the case of the $\alpha_1$
transition, and 4$b$ in the case of the $\alpha_2$ and $\alpha_3$
transitions.

In light of the above discussion, we suggest that the solution of
the long-standing discrepancies in $\alpha$-U between the Debye
temperatures obtained from calorimetry and elastic constant
measurements depends on the degree of pinning of the collective
mode. It is interesting to compare these energies for $\alpha$-U
with other well known CDW materials, see Fig.~\ref{Figure3}. At the
lowest energy, NbSe$_3$ represents the most widely studied CDW
material~\cite{Fuller}. Here, the single-particle energy gap was
studied as a function of defect concentration by subjecting the
samples to radiation damage. The threshold field increases linearly
with defect concentration supporting strong pinning similar to
TaSe$_3$~\cite{Hsei} and K$_{0.3}$MnO$_3$~\cite{Fuller}. In
$\alpha$-U the CDW state melts at T = 41~K ($\alpha_1$) giving a
frequency of 75~GHz for the collective mode. Here the doubling of
the unit cell in the $a$~direction cuts the Brillouin zone in half
forming a gap with energy 14.5~V/cm. Similarly, the $\alpha_2$
transition occurs by nesting of the Fermi surface with less density
of states in the $b$~direction. At $\alpha_3$ the collective mode is
confined to the bottom of the well. Hence, the pinning frequency and
the threshold electric field are higher.

\begin{figure}[t]
\includegraphics[width=\columnwidth]{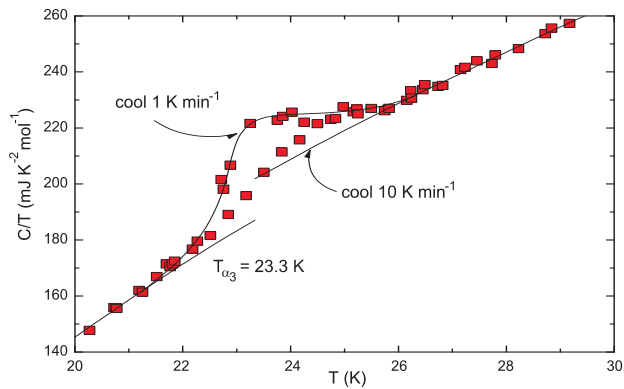}
\caption{(Color online) Dependence of the commensurate CDW
transition with cooling time. The transition is apparent only if the
sample has been cooled slowly (1K min$^{-1}$). With faster cooling
the $\alpha_3$ transition is no longer resolved, blending in with
the background.} \label{Figure4}
\end{figure}

The confinement of the collective modes to GHz frequencies in
$\alpha$-U is a result of topological defects combined with low
transformation temperatures. This effect is clear in the $\alpha_3$
transition where sluggish kinetics are observed, as shown in Fig. 4.
Upon slow cooling from the incommensurate state (35~K) to 20~K, and
then measuring on warming, the transition appears as a symmetric
bump. In contrast, when the sample is cooled 10 times faster, the
transition is not detected. Hysteretic effects are known to be more
pronounced with decreasing temperatures. As pointed out by Gr\"uner,
nonlinear I-V measurements in NbSe$_3$, the high temperature onset
is accompanied by oscillatory phenomena and evolves into a
hysteresis behavior at low temperature~\cite{Zettl}. This behavior
has been attributed to extended pinning centers or topological
defects~\cite{Brown}. The effect of slow time scales on the
collective dynamics of CDWs has been addressed by Myers and
Sethna~\cite{Myers}.

In summary, we have measured the specific heat difference between
single-crystal and polycrystalline $\alpha$-U and have described the
excess heat capacity as the contribution of the collective modes
above their respective pinning frequencies. Below the CDW melting
temperature of 41~K, the phonon splits into a mixture of acoustic-
and optic-like phonon modes as evidenced by the difference in Debye
temperatures, 256~K for the single crystal and 184~K for the
polycrystal. The long-standing discrepancy between Debye
temperatures obtained by calorimetry and elastic constants, can be
attributed to the degree of pinning of the collective mode. The
commensurate transition at 23~K, exhibits sluggish kinetics, likely
due to topological defects present at low temperatures. Similar
collective phenomena are likely to exist in other actinides.
Recently, a Kohn-like anomaly~\cite{Kohn} in the TA$_1$~[011]
branch, with softening of the [111] transverse modes, has been
observed in the vibrational spectrum of fcc-stabilized
plutonium~\cite{joe,one_more}, and the formation of the
$\alpha^{\prime}$-martensite lattice instability has been detected
in the low-temperature specific heat~\cite{myself}.

\begin{acknowledgments}
This work was carried out under the auspices of the U.S. Department
of Energy. The authors thank P.~S.~Riseborough, G.~H.~Lander and
M.~F.~Hundley for useful discussions.
\end{acknowledgments}


\begin{thebibliography}{99}
\bibitem{peierls_foot}  
Below the transition thermally activated carriers remain so that no
complete Peierls insulating state is formed.
\bibitem{excess_heat}   
In applying Eq.~\eqref{eq:cphi} to transition metal
chalcogenides~\cite{monceau} the excess specific heat , $C_\phi$,  was defined as
the difference between the specific heat of the sample and the Debye
specific heat.
\bibitem{lambda}     
Using the band structure $\gamma_{\mathrm{b.s.}}$ =
5.89~mJ~K$^2$~mol$^{-1}$~\cite{bandstructure}, the electron-phonon
parameter, $ \lambda =
  \raise0.2ex\hbox{$\gamma_{\mathrm{b.s.}}$} \!
  \mathord{\left/ {\vphantom{{\gamma_{b.s.} }\gamma}}\right.
  \kern-\nulldelimiterspace}\!\lower0.2ex\hbox{$\gamma $} - 1
  \>,
$
is determined to be 0.5 for the single crystal, and 0.7 for the
polycrystalline sample.
\bibitem{footnote}
   A direct experimental
   measurement of the pinning frequencies, $\nu_o$, and
   of the number of excitations,
   $N_\phi$, are necessary to validate the values quoted here.
\end{thebibliography}
\end{document}